\documentclass[preprint,nofootinbib,aps,superscriptaddress,eqsecnum]{revtex4-2} 
\usepackage{float}
\pdfoutput=1
\usepackage{extarrows}
\usepackage{tabularx}
\usepackage{graphicx}
\usepackage{amsmath}
\usepackage{amsfonts}
\usepackage{amssymb}
\usepackage{hyperref}
\usepackage[justification=centering]{caption} 
\usepackage{caption}
\usepackage{subcaption}
\usepackage{color}
\allowdisplaybreaks
\def\bea{\begin{eqnarray}}
	\def\eea{\end{eqnarray}}
\def\be{\begin{equation}}
	\def\ee{\end{equation}}

\begin{document}
	\title{Minimal Left-Right Symmetric Model with $A_4$ modular symmetry}
	\author{Ankita Kakoti}
	\email{ankitak@tezu.ernet.in}
	\author{Bichitra Bijay Boruah}
	\email{bijay@tezu.ernet.in}
	\author{Mrinal Kumar Das}
	\email{mkdas@tezu.ernet.in}
	\affiliation{Department of Physics, Tezpur University, Tezpur 784028, India}
	\begin{abstract}
		In this paper, we have realized the left-right symmetric model with modular symmetry. We have used $\Gamma$(3) modular group which is isomorphic to non-abelian discrete symmetry group $A_4$. The advantage of using modular symmetry is the non-requirement for the use of extra particles called 'flavons'. In this model, the Yukawa couplings are expressed in terms of modular forms $(Y_1,Y_2,Y_3)$. In this work, we have studied minimal Left-Right Symmetric Model for both type-I and type-II dominances. Here, we have calculated the values for the Yukawa couplings and then plotted it against the sum of the neutrino masses. The results obtained are well within the experimental limits for the desired values of sum of neutrino masses. We have also briefly analyzed the effects of the implications of modular symmetry on neutrinoless double beta decay with the new physics contributions within Left-Right Symmetric Model.
	\end{abstract}
\maketitle
\newpage
\begin{center}
	\section{\textbf{Introduction}}
\end{center}
\vspace{-0.4cm}
Despite the huge and continued success of the Standard Model (SM) of particle physics, it leaves some of the puzzles unanswered like the existence of neutrino masses, baryon asymmetry of the universe, existence of dark matter etc. The discovery of neutrino oscillation by Sudbury neutrino observatory and Super-Kamiokande experiments was a milestone discovery in the area of neutrino physics. The experiments like MINOS \cite{Evans:2013pka}, T2K \cite{T2K:2011ypd}, Daya-Bay \cite{DayaBay:2012fng}, Double-Chooz \cite{Suekane:2016ytt}, RENO \cite{Lasserre:2012ax} etc. provided evidence on the neutrinos being massive which is one of the most compelling revelation that we need to go beyond Standard Model. However inspite of the huge achievements in determining the neutrino oscillation parameters in solar, atmospheric , reactor and accelerator neutrino experiments, many questions related to neutrino still remain unsolved. Among these lies the question regarding the absolute mass scale of neutrinos, exact nature of the particle (Dirac or Majorana), hierarchical pattern of the mass spectrum (Normal or Inverted) and leptonic CP violation. The absolute mass scale of the neutrinos is not yet known. However experiments like Planck has given an upper bound on the sum of the light neutrino masses to be $\Sigma|m_{\nu_i}| <0.23 eV$ in 2012 \cite{Planck:2013pxb} and recently the bound has been constarined to $\Sigma|m_{\nu_i}| <0.11 eV$ \cite{Planck:2018vyg}. The most successful data pertaining to neutrino oscillation parameters is found in the 3$\sigma$ global fit data \cite{Esteban:2020cvm} as shown in table (1) .
\begin{table}[H]
	\begin{center}
		\begin{tabularx}{0.5\textwidth}{| >{\raggedright\arraybackslash}X
				| >{\centering\arraybackslash}X
				| >{\raggedright\arraybackslash}X |}
			\hline
			Parameters & Normal Ordering & Inverted Ordering \\
			\hline
			$\Delta$ $m_{21}^2$ $(10^{-5}eV^2)$ & $6.82$ $\rightarrow$ $8.04$ & $6.82$ $\rightarrow$ $8.04$\\ 
			\hline
			$\Delta$ $m_{3l}^2$ $(10^{-5}eV^2)$ & $2.435$ $\rightarrow$ $2.598$ & $-2.581$ $\rightarrow$ $-2.414$ \\ 
			\hline
			$sin^2$ $\theta_{12}$ & $0.264$ $\rightarrow$ $0.343$ & $0.269$ $\rightarrow$ $0.343$ \\
			\hline
			$sin^2$ $\theta_{23}$ & $0.415$ $\rightarrow$ $0.616$ & $0.419$ $\rightarrow$ $0.617$ \\
			\hline
			$sin^2$ $\theta_{13}$ & $0.02032$ $\rightarrow$ $0.02410$ & $0.02052$ $\rightarrow$ $0.02428$ \\
			\hline
		\end{tabularx}
	\end{center}
	\caption{\label{tab:Table 1}Global fit 3$\sigma$ values for neutrino oscillation parameters.}
\end{table}
 We have used the definition,\\
\begin{equation}
	\label{E:first}
	\Delta m_{3l}^2 = \Delta m_{31}^2 ; \Delta m_{31}^2 >0 ; NO
\end{equation}\\
\vspace{-1.5cm}

\begin{equation}
	\label{E:second}
	\Delta m_{3l}^2 = \Delta m_{32}^2 ; \Delta m_{32}^2<0 ; IO
\end{equation}

The simplest way to look for neutrino masses is by the seesaw mechanism. The mechanism may be of type I \cite{Schechter:1980gr}, \cite{Mohapatra:1979ia},type II \cite{Wetterich:1981bx}, \cite{Antusch:2004xy},type III \cite{Foot:1988aq} and Inverse Seesaw \cite{Haba:2016lxc}. These are extensions of the SM where we incorporate extra particles like right-handed fermions,scalar fermion triplets, gauge singlet neutral fermions etc. The BSM physics also sheds light upon the phenomena like baryon asymmetry of the universe (BAU) \cite{Kolb:1979ui}, Lepton Number Violation (LNV) \cite{Bilenky:2012qi}, Lepton Flavor violation (LFV) \cite{Altarelli:2012bn}, existence of dark matter \cite{Taoso:2007qk}, \cite{Turner:1993dra} etc. A BSM framework which has been successful in explaining the first three of the phenomenologies is the Left- Right Symmetric Model (LRSM) \cite{Corrigan:2015kfu,Pati:1974yy,Mohapatra:1974gc,BhupalDev:2018xya,Senjanovic:1979wr}, an extension of the SM corresponding to the addition of $SU(2)_R$ group into the theory. The gauge group of LRSM is  $SU(3)_C\otimes SU(2)_R\otimes SU(2)_L\otimes U(1)_{B-L}$. The type I and type II seesaw masses appear naturally in the model. The right-handed neutrinos are an essential part of the model, which acquires Majorana mass when $SU(2)_R$ symmetry is broken. LRSM provides a natural framework to understand the spontaneous breaking of parity and origin of small neutrino masses by seesaw mechanism \cite{Grimus:1993fx}.\\
Another concerning aspect is the ambiguity regarding nature of neutrinos which has not been yet predicted by the SM of particle physics, that whether neutrinos are Dirac or Majorana fermions. This problem is directly connected to the issue of lepton number conservation. One of the process of fundamental importance which arises in almost any extension of the SM is Neutrinoless Double Beta Decay(NDBD) \cite{Vergados:2012xy}, \cite{Cardani:2018lje} which when verified can assure that neutrinos are Majorana fermions. NDBD is a slow, radiative process that transforms a nuclide of atomic number Z into its isobar with atomic number Z+2 \cite{Boruah:2021ktk},
\begin{equation}
	\label{E:third}
	N(A,Z) \rightarrow N(A,Z+2)+e^- +e^-     
\end{equation}
The main aim in the search of NDBD (0$\nu\beta\beta$) is the measurement of effective Majorana neutrino mass, which is a combination of the neutrino mass eigenstates and neutrino mixing matrix terms \cite{Boruah:2021ktk}. However, no experimental evidence regarding the decay has been in picture till date. In addition to the determination of the effective masses, the half-life of the decay \cite{Ge:2017erv} combined with sufficient knowledge of the nuclear matrix elements (NME), we can set a constraint on the neutrino masses. The experiments like KamLAND-Zen \cite{Shirai:2018ycl} and GERDA \cite{GERDA:2020xhi} which uses Xenon-136 and Germanium-76 respectively have improved the lower bound on the half-life of the decay process. However, KamLAND-Zen imposes the best lower limit on the half life as $T_{1/2}^{0\nu}>1.07\times10^{26}$ yr at 90 \% CL and the corresponding upper limit of the effective Majorana mass in the range (0.061-0.165)eV. There are several contributions in LRSM that appear due to additional RH current interactions, giving rise to sizeable LFV rates for TeV scale RH neutrino that occur at rates accessible in current experiments. It has been found that the most significant constraints has been provided by the decays, $\mu\rightarrow3e$ and $\mu\rightarrow\gamma e$. In the Standard Model, these LFV decays are suppressed by the tiny neutrino masses. No experiment has so far observed any flavor violating processes including charged leptons. However, many experiments are currently going on to set strong limits on the most relevant LFV observables that will constrain the parameter space of many new models. The best bounds on the branching ratio for LFV decays of the form $\mu\rightarrow\gamma e$ comes from MEG experiment and it is set at $BR(\mu\rightarrow\gamma e)< 4.2 \times 10^{-13}$. In case of the decay $\mu\rightarrow 3e$, the bound is set by the SINDRUM experiment at $BR(\mu\rightarrow 3e)< 1.0 \times 10^{-12}$.  \par
As mentioned LRSM is an important theory that incorporates the above mentioned phenomenologies, i.e., the phenomenologies related to neutrinos. There are many works where the authors make use of discrete symmetry groups like $A_4$ \cite{Ma:2001dn},$S_4$ \cite{Altarelli:2009gn},$Z_2$ etc. \cite{Altarelli:2010gt} to analyze the problem of flavor structure of fermions and to study various related phenomenologies. In our work, we have used $A_4$ modular symmetry to study neutrino masses and mixings and hence study Neutrinoless Double Beta Decay within the model. The advantage of using modular symmetry over discrete flavor symmetries is that the study of the model using symmetries can be done without the introduction of extra particles called 'flavons'. Hence the model is minimal. \par
However, in this work we have not done a very detailed analysis of the above mentioned phenomenologies, but only realized the left-right symmetric model with the help of $A_4$ modular symmetry and studied the variations of new physics contributions of neutrinoless double beta decay within LRSM with the range of values for Yukawa couplings, which in our model is expressed as modular forms. In section (II), we have given a detailed explanation of the left-right symmetric model, the associated Lagrangian and the mass terms. We begin section (III) by introducing 
\newpage
modular symmetry and then in section (IV), we incorporate modular symmetry into LRSM and determine the associated mass matrices. In section (V), we present a very brief discussion of neutrinoless double beta decay and its associated contributions and their relations with the modular forms. In section (VI), the numerical analysis and results of this work has been discussed and the last section reads the conclusion for the present work.

\begin{center}
	\section{\textbf{Minimal Left-Right Symmetric Model}}
\end{center}

	The Left-Right Symmetric Model (LRSM) was first introduced around 1974 by Pati and Salam. Rabindra N. Mohapatra and Goran Senjanovic were also some pioneers of this very elegant theory. LRSM is an extension of the Standard Model of particle physics, the gauge group being $SU(3)_C\otimes SU(2)_R\otimes SU(2)_L\otimes U(1)_{B-L}$, which has been studied by several groups since 1970's  \cite{Grimus:1993fx}, \cite{Pati:1974yy, Mohapatra:1974gc,BhupalDev:2018xya,Senjanovic:1979wr}. The usual type-I and type-II seesaw neutrino masses arises naturally in the model. The seesaw scale is identified by the breaking of $SU(2)_R$. Some other problems are also addressed in LRSM like parity, CP violation of weak interaction, massive neutrinos, hierarchy problems, etc. LRSM removes the disparity between the left and right-handed fields by considering the RH fields to be doublet under the additional $SU(2)_R$ keeping the right sector couplings same as the left-one by left-right symmetry. In this model, the electric charge is given by $Q=T_{3L}+T_{3R}+\frac{B-L}{2}$, where $T_{3L}$ and $T_{3R}$ are the generators of $SU(2)_L$ and $SU(2)_R$ respectively. $B-L$ refers to baryon number minus lepton number. The particle content of the model along with their respective charge assignments are given in table(III).
 The matrix representation for the scalar sector is given by,
\begin{equation}
	\label{E:fourth}
	\phi = \begin{pmatrix}
		\phi_1^0 & \phi_1^+\\
		\phi_2^- & \phi_2^0
	\end{pmatrix}
\end{equation}
\begin{equation}
	\label{E:fifth}
	\Delta_{L,R} = \begin{pmatrix}
		\frac{\delta_{L,R}^+}{\sqrt{2}} & \delta_{L,R}^{++}\\
		\delta_{L,R}^0 & -\frac{\delta_{L,R}^+}{\sqrt{2}}
	\end{pmatrix}
\end{equation}
In order for the fermions to attain mass, a Yukawa Lagrangian is necessary which couples to the bidoublet $\phi$. The Yukawa Lagrangian incorporating the bidoublet is given by,
 \begin{equation}
 	\label{E:sixth}
	\mathcal{L_{D}} = \overline{l_{iL}}(Y_{ij}^l \phi + \widetilde{Y_{ij}^l}\widetilde{\phi})l_{jR} + \overline{Q_{iL}}(Y_{ij}^q \phi + \widetilde{Y_{ij}^q}\widetilde{\phi})Q_{jR} + h.c   
\end{equation}
\newpage
where, $l_L$ and $l_R$ are the left-handed and right-handed lepton fields, $Q_L$ and $Q_R$ are the left-handed and right-handed quark fields. $Y^l$ being the Yukawa coupling corresponding to leptons and $Y^q$ being the Yukawa coupling for the quarks. 
The Yukawa Lagrangian incorporating the scalar triplets which play a role in providing Majorana mass to the neutrinos is given by,
\begin{equation}
	\label{E:7}
	\mathcal{L_M}=f_{L,ij}{\Psi_{L,i}}^TCi\sigma_2\Delta_L\Psi_{L,j}+f_{R,ij}{\Psi_{R,i}}^TCi\sigma_2\Delta_R\Psi_{R,j}+h.c
\end{equation}
$f_L$ and $f_R$ are the Majorana Yukawa couplings and are equal subjected to discrete left-right symmetry. The scalar potential in LRSM is a combination of interaction terms consisting the potential and after spontaneous symmetry breaking the scalars attain VEVs given by,
 \begin{equation}
 	\label{E:8}
 	<\Delta_{L,R}> = \frac{1}{\sqrt{2}} \begin{pmatrix}
 		0 & 0 \\
 		v_{L,R} & 0
 	\end{pmatrix}
 \end{equation}
\begin{equation}
	\label{E:9}
	<\phi> = \begin{pmatrix}
		k & 0 \\
		0 & e^{i\theta} k'
	\end{pmatrix}
\end{equation}
The magnitudes of the VEVs follows the relation, $|v_L|^2 < |k^{2} +k'^{2}| < |v_R|^2$. The breaking pattern of the LRSM gauge group takes place in two steps. The LRSM gauge group is first broken down to the Standard Model gauge group by the vev of the scalar triplet $\Delta_R$, and then the Standard Model gauge group is broken down to the electromagnetic gauge group i.e., $U(1)_{em}$ by the vev of the bidoublet and a tiny vev of the scalar triplet $\Delta_L$.\\
The Dirac mass terms for the leptons come from the Yukawa Lagrangian, which for the charged leptons and the neutrinos are given by,
\begin{equation}
	\label{E:10}
	M_l = \frac{1}{\sqrt{2}}(k' Y_l + k \tilde{Y_l})
\end{equation}
\begin{equation}
	\label{E:11}
	M_D = \frac{1}{\sqrt{2}}(k Y_l + k' \tilde{Y_l})
\end{equation}

 The light neutrino mass after spontaneous symmetry breaking (SSB), generated within a type (I+II) seesaw can be written as,

\begin{equation}
	\label{E:12}
	M_\nu= {M_\nu}^{I}+{M_\nu}^{II},
\end{equation}

\begin{equation}
	\label{E:13}
	M_\nu=M_D{M_{RR}}^{-1}{M_D}^T + M_{LL}
\end{equation}
where, 
\begin{equation}
	\label{E:14}
	M_{LL} = \sqrt{2} v_L f_L
\end{equation}
and, \begin{equation}
	\label{E:15}
	M_{RR} = \sqrt{2} v_R f_R
\end{equation}
The first and second terms in \label{eq} corresponds to type-I seesaw and type-II seesaw masses respectively. It is an interesting fact that in the context of LRSM both type-I and type-II terms can be equally dominant or either of the two terms can be dominant, but under certain conditions \cite{Borah:2013bza,Luo:2008rs}. It has been demonstrated in the Appendix A. 
In the context of LRSM however, both the type-I and type-II mass terms can be expressed in terms of the heavy right-handed Majorana mass matrix, so equation \eqref{E:13} will follow,
\begin{equation}
	\label{E:16}
	M_\nu = M_D M_{RR}^{-1} M_D^T + \gamma\Biggl(\frac{M_W}{v_R}\Biggl)^2 M_{RR}
\end{equation}
where, $\gamma$ is a dimensionless parameter which is a function of various couplings, appearing in the VEV of the triplet Higgs $\Delta_L$, i.e., $v_L = \gamma (\frac{v^2}{v_R})$ and here, $v = \sqrt{k^2 + k'^2}$, and
\begin{equation}
	\label{E:17}
	\gamma =  \frac{\beta_1 k k' + \beta_2 k^2 + \beta_3 k'^2}{(2\rho_1 - \rho_3)(k^2 + k'^2)}
\end{equation} 
In our model, the dimensionless parameter $\gamma$ has been fine tuned to $\gamma \approx 10^{-6}$ and $v_R$ is of the order of $TeV$.

\begin{center}
	\section{\textbf{Modular Symmetry}}
\end{center}

Modular symmetry has gained much importance in aspects of model building \cite{King:2020qaj}, \cite{Novichkov:2019sqv}. This is because it can minimize the extra particle called 'flavons' while analyzing a model with respect to a particular symmetry group. An element $q$ of the modular group acts on a complex variable $\tau$ which belongs to the upper-half of the complex plane given as \cite{Novichkov:2019sqv} \cite{Feruglio:2017spp}  \par
\begin{equation}
	\label{E:18}
	q\tau = \frac{a\tau+b}{c\tau+d}
\end{equation}
where $a,b,c,d$ are integers and $ad-bc=1$, Im$\tau$$>$0.\par
The modular group is isomorphic to the projective special linear group PSL(2,Z) = SL(2,Z)/$Z_2$ where, SL(2,Z) is the special linear group of integer $2\times2$ matrices having determinant unity and $Z_2=({I,-I})$ is the centre, $I$ being the identity element. The modular group can be represented in terms of two generators $S$ and $T$ which satisfies $S^2=(ST)^3=I$. $S$ and $T$ satisfies the following matrix representations:
\begin{equation}
	\label{E:19}
	S = \begin{pmatrix}
		0 & 1\\
		-1 & 0
	\end{pmatrix}
\end{equation}
\begin{equation}
	\label{e:20}
	T = \begin{pmatrix}
		1 & 1\\
		0 & 1
	\end{pmatrix}
\end{equation}
corresponding to the transformations,\\
\begin{equation}
	\label{e:21}
	S : \tau \rightarrow -\frac{1}{\tau} ; T : \tau \rightarrow \tau + 1
\end{equation}

Finite modular groups (N $\le$ 5) are isomorphic to non-abelian discrete groups, for example,
$\Gamma(3) \approx A_4$, $\Gamma(2) \approx S_3$, $\Gamma(4) \approx S_4$. While using modular symmetry, the Yukawa couplings can be expressed in terms of modular forms, and the number of modular forms present depends upon the level and weight of the modular form. For a modular form of level N and weight 2k, the table below shows the number of modular forms associated within and the non-abelian discrete symmetry group to which it is isomorphic \cite{Feruglio:2017spp}.
\begin{table}[H]
	\begin{center}
		\begin{tabular}{|c|c|c|}
			\hline
			N & No. of modular forms & $\Gamma(N)$ \\
			\hline
			2 & k + 1 & $S_3$ \\
			\hline
			3 & 2k + 1 & $A_4$ \\
			\hline
			4 & 4k + 1 & $S_4$ \\
			\hline
			5 & 10k + 1 & $A_5$ \\
			\hline 
			6 & 12k &  \\
			\hline
			7 & 28k - 2 & \\
			\hline
		\end{tabular}
		\caption{No. of modular forms corresponding to modular weight 2k.}
	\end{center}
\end{table}
\newpage
In our work, we will be using modular form of level 3, that is, $\Gamma(3)$ which is isomorphic to $A_4$ discrete symmetry group. The weight of the modular form is taken to be 2, and hence it will have three modular forms $(Y_1,Y_2,Y_3)$ which can be expressed as expansions of q given by,
\begin{equation}
	\label{e:22}
	Y_1 = 1 + 12 q + 36 q^2 + 12 q^3 + 84 q^4 + 72 q^5 + 36 q^6 + 96 q^7 + 
	180 q^8 + 12 q^9 + 216 q^{10}
\end{equation}
\begin{equation}
	\label{e:23}
	Y_2 = -6 q^{1/3} (1 + 7 q + 8 q^2 + 18 q^3 + 14 q^4 + 31 q^5 + 20 q^6 + 
	36 q^7 + 31 q^8 + 56 q^9)
\end{equation}
\begin{equation}
	\label{e:24}
	Y_3 = -18 q^{2/3} (1 + 2 q + 5 q^2 + 4 q^3 + 8 q^4 + 6 q^5 + 14 q^6 + 
	8 q^7 + 14 q^8 + 10 q^9)
\end{equation}
where, $q = \exp(2\pi i \tau)$.

\begin{center}
	\section{Minimal LRSM with $A_4$ modular symmetry}
\end{center}
In particle physics, symmetries have always played a very crucial role. The realization of LRSM with the help of discrete flavor symmetries have been done in earlier works \cite{Duka:1999uc}, \cite{Boruah:2022bvf}. In our work we have incorporated $A_4$ modular symmetry into LRSM. The advantage of using modular symmetry rather than flavor symmetry is the minimal use of extra particles (flavons) and hence the model is minimal. The model contains usual particle content of LRSM \cite{Sahu:2020tqe}. The lepton doublets transform as triplets under $A_4$ and the bidoublet and scalar triplets transform as 1 under $A_4$ \cite{Abbas:2020qzc}. As we have considered modular symmetry, we assign modular weights to the particles, keeping in mind that matter multiplets corresponding to the model can have negative modular weights, but the modular forms cannot be assigned negative weights. The assignment of these weights are done in such a way that in the Lagrangian the sum of the modular weights in each term is zero. Modular weights corresponding to each particle is shown in table (III). The Yukawa Lagrangian for the leptonic and quark sector in LRSM is given by equation \eqref{E:sixth},\eqref{E:7} and with a reference to that we can write the Yukawa Lagrangian of our $A_4$ modular symmetric LRSM, for the fermionic sector, by introducing Yukawa coupling in the form of modular forms $Y$ is given as,
\begin{equation}
	\label{e:25}
	\mathcal{L_Y} = \overline{l_L}\phi{l_R}Y+\overline{l_L}\tilde{\phi}{l_R}Y+\overline{Q_L}\phi{Q_R}Y+\overline{Q_L}\tilde{\phi}{Q_R}Y+{{l_R}^T}C i{\tau_2}{\Delta_R}{l_R}{Y}+{{l_L}^T}C i{\tau_2}{\Delta_L}{l_L}{Y}
\end{equation}
\newpage
\begin{table}[H]
	\begin{center}
		\begin{tabular}{|c|c|c|c|c|c|c|c|c|}
			\hline
			Gauge group & $Q_L$ & $Q_R$ & $l_L$ & $l_R$ & $\phi$ & $\Delta_L$ & $\Delta_R$ \\
			\hline
			$SU(3)_C$ & 3 & 3 & 1 & 1 & 1 & 1 & 1\\
			\hline
			$SU(2)_L$ & 2 & 1 & 2 & 1 & 2 & 3 & 1 \\
			\hline
			$SU(2)_R$ & 1 & 2 & 1 & 2 & 2 & 1 & 3 \\
			\hline
			$U(1)_{B-L}$ & 1/3 & 1/3  & -1 & -1 & 0 & 2 & 2 \\
			\hline
			$A_4$ & 3 & 3 & 3 & 3 & 1 & 1 & 1 \\
			\hline 
			$k_I$ & 0 & -2 & 0 & -2 & 0 & -2 & 2 \\
			\hline
		\end{tabular}
		\caption{\label{tab:Table 2}Charge assignments for the particle content of the model.}
	\end{center}
\end{table}

The Yukawa couplings $Y = (Y_1,Y_2,Y_3)$ are expressed as modular forms of level 3.
\begin{table}[H]
	\begin{center}
		\begin{tabular}{|c|c|}
			\hline
			& Y (modular forms)\\
			\hline
			$A_4$ & $3$ \\
			\hline
			$k_I$ & $2$ \\
			\hline
		\end{tabular}
		\caption{\label{tab:Table 3}Charge assignment and modular weight for the corresponding modular Yukawa form for the model.}
	\end{center}
\end{table}
In our work, we are concerned with the mass of the neutrinos and as such, using $A_4$ modular symmetry and using the multiplication rules for $A_4$ group, we construct the Dirac and Majorana mass matrices as given below. The Dirac mass matrix is given by,
\begin{equation}
	\label{e:26}
	M_D =v\begin{pmatrix}
		2Y_1 & -Y_3 & -Y_2\\
		-Y_2 & -Y_1 & 2Y_3\\
		-Y_3 & 2Y_2 & -Y_1
	\end{pmatrix}
\end{equation}
where, $v$ is considered to be the VEV for the Higgs bidoublet.

The right-handed Majorana mass matrix is given by,
\begin{equation}
	\label{e:27}
	M_R =v_R\begin{pmatrix}
		2Y_1 & -Y_3 & -Y_2\\
		-Y_3 & 2Y_2 & -Y_1\\
		-Y_2 & -Y_1 & 2Y_3
	\end{pmatrix}
\end{equation}
where, $v_R$ is the VEV for the scalar triplet $\Delta_R$. As it is seen that the Majorana mass matrix for our model is found to be symmetric in nature as it should be. Under these assumptions for modular symmetric LRSM and the basis that we have considered, our charged lepton mass matrix is also found to be diagonal.

The type-I seesaw mass is then given by,
\begin{equation}
	\label{e:28}
	M_{\nu_I}=M_D.{M_R}^{-1}.{M_D}^T 
\end{equation}
and, the type-II seesaw mass is given by,
\begin{equation}
	\label{e:29}
	M_{\nu_{II}}=M_{LL}
\end{equation}
As mentioned above, in LRSM type-II seesaw mass can also be expressed in terms of the right-handed mass $M_R$ as,
\begin{equation}
	\label{e:30}
	M_{\nu_{II}}=\gamma \Biggl(\frac{M_W}{v_R}\Biggl)^2 M_R
\end{equation}
\subsection{Type-I dominanace}
In LRSM, the type-I seesaw mass dominates when the vev of the left-handed triplet is taken to be negligibly small and hence the type-II term is absent. In such a case the lightest neutrino mass can be given in terms of the type-I seesaw mass term given by,
\begin{equation}
	\label{e:31}
	M_{\nu}=M_D{M_{R}}^{-1}{M_D}^T 
\end{equation}
and the heavy right-handed Majorana mass term can be given as,
\begin{equation}
	\label{e:32}
	M_{R}=f_R v_R
\end{equation}
where, $f_R$ is the right-handed Majorana Yukawa coupling. \\
In the approximation that $k'<< k$, and if we consider that our Yukawa coupling $Y^l$ corresponding to the neutrino masses is $y_D$ and the coupling $\widetilde{Y^l}$ for the charged fermion masses is denoted by $y_L$, so considering $y_D k >> y_L k'$  we can write the type-I mass term as \cite{Chakrabortty:2012mh},
\begin{equation}
	\label{e:33}
	M_\nu = \frac{k^2}{v_R} y_D f_R^{-1} y_D^T 
\end{equation}
If we consider that $U_R$ is a unitary matrix that diagonalizes $M_R$, so since the VEV $v_R$ is a constant the same matrix will also diagonalize the coupling matrix $f_R$. Taking $f_R = f_L =f$, so
\begin{equation}
	\label{e:34}
	f = U_R f^{dia} U_R^T
\end{equation}
If we take inverse on both sides and taking into account the property of a unitary matrix $(U_R^{-1} = U_R^T)$, we get,
\begin{equation}
	\label{e:35}
	f^{-1} = U_R^T(f^{dia})^{-1}U_R
\end{equation}
Therefore, we get
\begin{equation}
	\label{e:36}
	M_\nu = \frac{k^2}{v_R} y_D U_R^T (f^{dia})^{-1} U_R y_D^T
\end{equation}
Multiplying both sides of the equation with $U_R^T$ from the right and with $U_R$ from left, we finally arrive at the following equation,
\begin{equation}
	\label{e:37}
	U_R M_\nu U_R^T = (M_\nu)^{dia}
\end{equation}
where we have used $U_R y_D U_R^T = y_D$. So, the unitary matrix diagonalizing the matrix $M_R$ also diagonalizes the light neutrino mass matrix. So in this case it can be determined that if $m_i$ denotes the light neutrino mass and $M_i$ denotes the heavy neutrino mass, then they are related as 
\begin{equation}
	\label{e:38}
	m_i \propto \frac{1}{M_i}
\end{equation}
For our model, the Yukawa couplings are modular forms expressed as expansions of $q$, and the mass matrices are expressed in terms of the modular forms $(Y_1,Y_2,Y_3)$. So, the light neutrino mass matrix, $M_{\nu}$ for type-I dominance is given by the equation \eqref{e:31}. As already stated in equations \eqref{e:26} and \eqref{e:27}, the Dirac and Majorana mass matrices are determined by the application of multiplication rules for the $A_4$ group. So, for type-I dominance, our light neutrino mass matrix will be given by,
\begin{equation}
	\label{e:39}
	M_{\nu}=\frac{v^2}{v_R}\begin{pmatrix}
		2Y_1 & -Y_2 & -Y_3 \\
		-Y_2 & 2Y_3 & -Y_1 \\
		-Y_3 & -Y_1 & 2Y_2
	\end{pmatrix}
\end{equation}
As mentioned previously, the value for $v_R$ is of the order of $TeV$ and that for $v$ is in $GeV$. We have computed the values of the sum of the neutrino masses for type-I dominance and checked the correctness of our model by plotting it against the Yukawa couplings and the result was found to match the experimental bounds.
\begin{figure}[H]
	\centering
\captionsetup{justification=centering}
\includegraphics[scale=0.458]{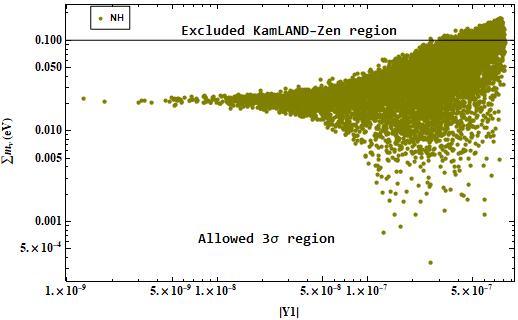}
\includegraphics[scale=0.4]{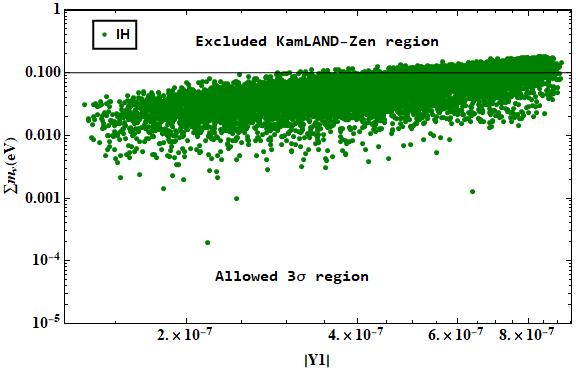}\\
\caption{Variation of $|Y_1|$ with sum of neutrino masses.}
\end{figure}
\begin{figure}[H]
	\centering
	\captionsetup{justification=centering}
	\includegraphics[scale=0.458]{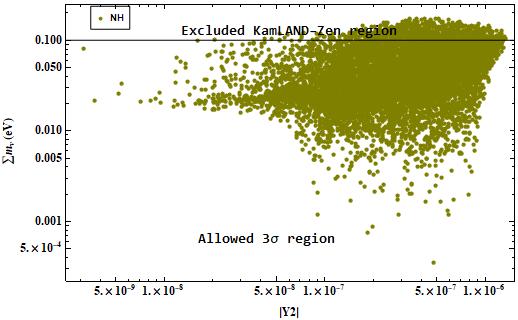}
	\includegraphics[scale=0.4]{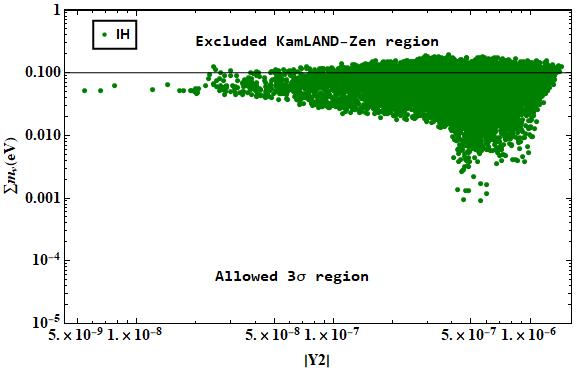}\\
	\caption{Variation of $|Y_2|$ with sum of neutrino masses.}
\end{figure}

\begin{figure}[H]
	\centering
	\captionsetup{justification=centering}
	\includegraphics[scale=0.458]{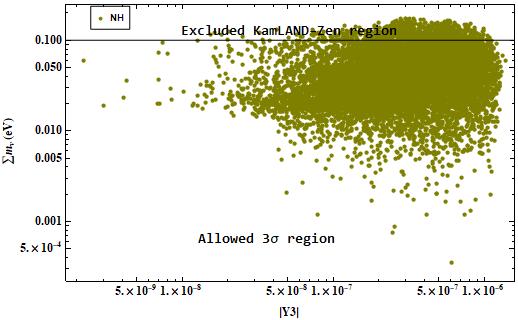}
	\includegraphics[scale=0.4]{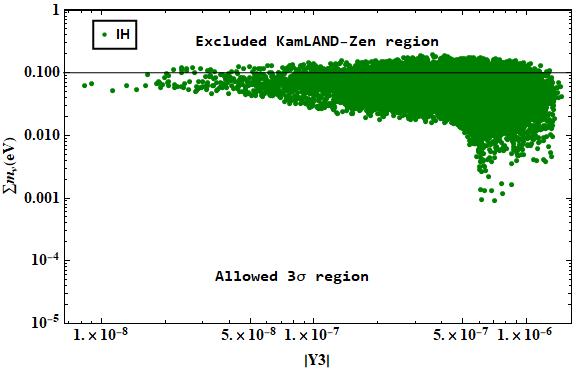}\\
	\caption{Variation of $|Y_3|$ with sum of neutrino masses.}
\end{figure}

\subsection{Type-II dominance}
Type-II seesaw mass in LRSM dominates when the Dirac term connecting the right-handed and left-handed parts is negligible as compared to that of the type-II term \cite{Chakrabortty:2012mh}. In that case, our light neutrino mass $m_\nu$ will given by the type-II seesaw mass term, i.e.,
\begin{equation}
	\label{e:40}
	M_{\nu_L} = f_L v_L
\end{equation}
And the heavy mass matrix is given by,
\begin{equation}
	\label{e:41}
	M_R = f_R v_R
\end{equation}
Again if we consider that $U_L$ and $U_R$ diagonalizes $M_{\nu_L}$ and $M_R$ respectively, so for the reason mentioned above the same matrices will also diagonalize $f_L$ and $f_R$ respectively and since in our model, $f_L = f_R$, so we can consider $U_L = U_R$. In such a case, we arrive at an important result that
\begin{equation}
	\label{e:42}
	m_i \propto M_i
\end{equation}
Now using modular symmetry the light neutrino mass matrix for type-II dominance in our model is given by,
\begin{equation}
	\label{e:43}
	m_\nu = v_L \begin{pmatrix}
		2Y_1 & -Y_3 & -Y_2 \\
		-Y_3 & 2Y_2 & -Y_1 \\
		-Y_2 & -Y_1 & 2Y_3
	\end{pmatrix}
\end{equation}
where, $v_L$ is the vev for left-handed scalar triplet.
The value of $v_L$ is taken to be of the order of $eV$. The sum of the neutrino masses is computed for type-II dominance and plotting of the sum is done with the Yukawa couplings which are found to be as shown under,
\begin{figure}[H]
	\centering
	\captionsetup{justification=centering}
	\includegraphics[scale=0.4]{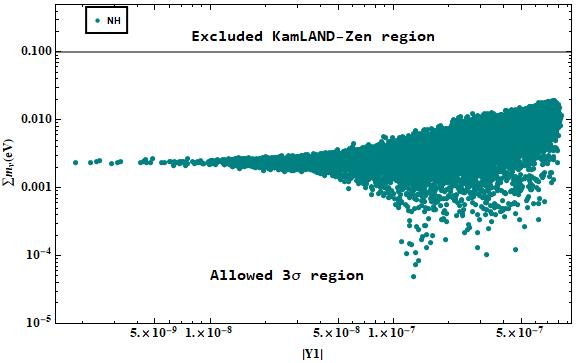}
	\includegraphics[scale=0.4]{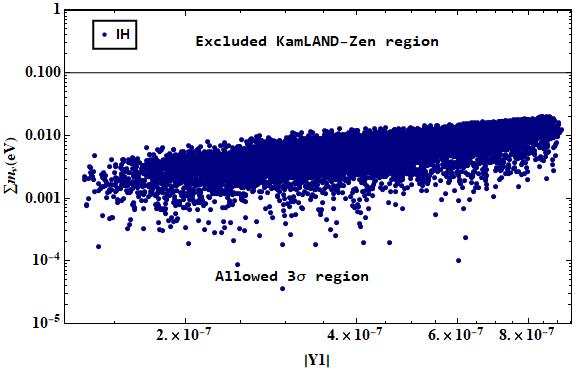}\\
	\caption{Variation of $|Y_1|$ with sum of neutrino masses.}
\end{figure}
\begin{figure}[H]
	\centering
	\captionsetup{justification=centering}
	\includegraphics[scale=0.4]{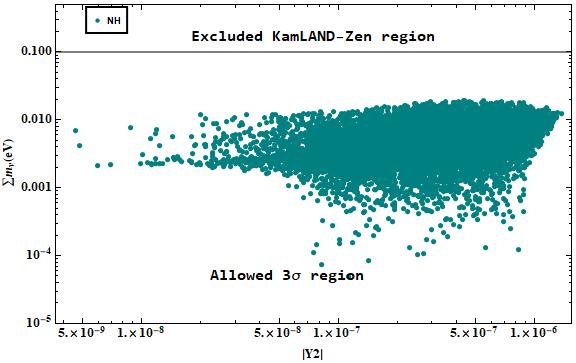}
	\includegraphics[scale=0.4]{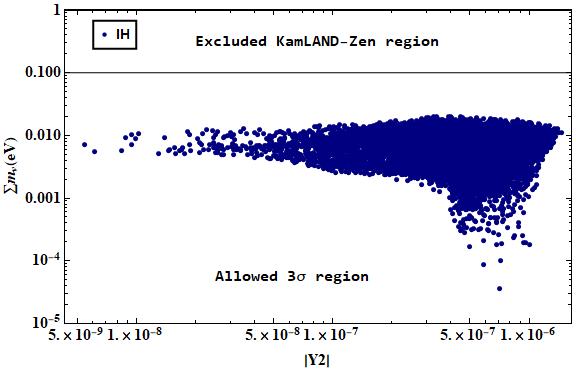}\\
	\caption{Variation of $|Y_2|$ with sum of neutrino masses.}
\end{figure}
\begin{figure}[H]
	\centering
	\captionsetup{justification=centering}
	\includegraphics[scale=0.4]{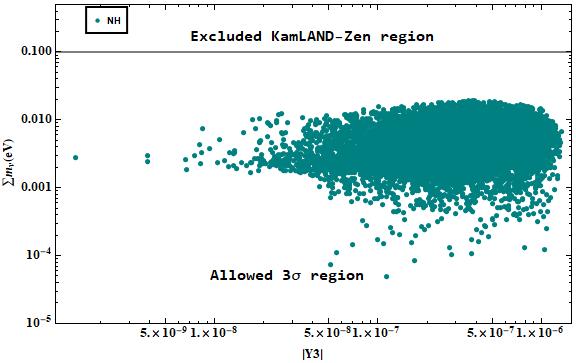}
	\includegraphics[scale=0.4]{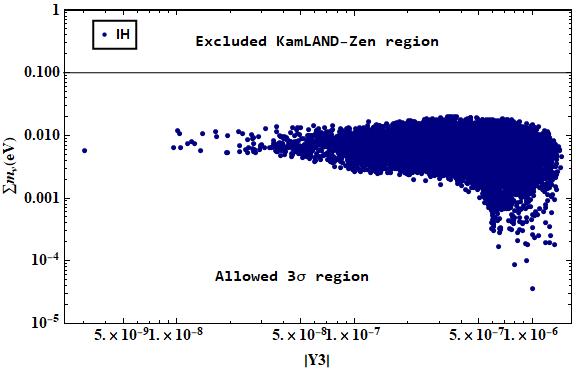}\\
	\caption{Variation of $|Y_3|$ with sum of neutrino masses.}
\end{figure}

\begin{center}
	\section{Neutrinoless Double Beta Decay $(0\nu\beta\beta)$ in minimal LRSM}
\end{center}
Neutrinoless double beta decay is a lepton number violating process, which if proven to exist will directly imply the Majorana nature of neutrinos.
\begin{equation}
	\label{e:44}
	N(A,Z) \rightarrow N(A,Z+2)+e^{-}+e^{-}
\end{equation}
Many groups have however already done a lot of work on NDBD in the model , \cite{Pati:1974yy},\cite{Gu:2015uek,Barry:2012ga,Li:2020flq,Borgohain:2017inp,Ge:2015yqa,Deppisch:2017vne,Boruah:2021ktk}. In LRSM \cite{Senjanovic:1975rk}, there are several contributions to NDBD in addition to the standard contribution via light Majorana neutrino exchange owing to the presence of several heavy additional scalar,vector and fermionic fields \cite{Awasthi:2016kbk,Bambhaniya:2015ipg,Borah:2010zq,Borgohain:2017akh}. Various contributions to NDBD transition rate in LRSM are discussed as follows :
\begin{itemize}
	\item Standard Model contribution to NDBD where the intermediate particles are the $W_L$ bosons and light neutrinos, the process in which the amplitude depends upon the leptonic mixing matrix elements and light neutrino masses.
	\item Heavy right-handed neutrino contribution in which the mediator particles are the $W_L$ bosons and the amplitude depends upon the mixing between light and heavy neutrinos as well as the mass of the heavy neutrino.
	\item  Light neutrino contribution to NDBD where the intermediate particles are $W_R$ bosons and the amplitude depends upon the mixing between light and heavy neutrinos as well as mass of the right-handed gauge boson $W_R$.
	\item  Heavy right-handed neutrino contribution where the mediator particles are the $W_R$ bosons. The amplitude of this process is dependent on the elements of the right handed leptonic mixing matrix and mass of the right-handed gauge boson, $W_R$ as well as the mass of the heavy right handed Majorana neutrino.
	\item  Light neutrino contribution from the Feynman diagram mediated by both $W_L$ and $W_R$, and the amplitude of the process depends upon the mixing between light and heavy neutrinos,leptonic mixing matrix elements, light neutrino masses and the mass of the gauge bosons, $W_L$ and $W_R$. 
	\item Heavy neutrino contribution from the Feynman diagram mediated by both $W_L$ and $W_R$, and the amplitude of the process depends upon the right handed leptonic mixing matrix elements, mixing between the light and heavy neutrinos, also the mass of the gauge bosons, $W_L$ and $W_R$ and the mass of the heavy right handed neutrino.
	\item Scalar triplet contribution ($\Delta_L$) in which the mediator particles are $W_L$ bosons, and the amplitude for the process depends upon the masses of the $W_L$ bosons, left-handed triplet Higgs, as well as their coupling to leptons.
	\item Right-handed scalar triplet contribution ($\Delta_R$) contribution to NDBD in which the mediator particles are $W_R$ bosons, and the amplitude for the process depends upon the masses of the $W_R$ bosons, right-handed triplet Higgs, $\Delta_R$ as well as their coupling to leptons. 
\end{itemize}
In our work, where we have incorporated $A_4$ modular symmetry to LRSM and in our present work we have considered three of the above mentioned contributions, one from the standard light neutrino contribution and the other two new physics contribution mediated by $W_R^{-}$ and $\Delta_R$ respectively. For simple approximations, an assumption has been made in the mass scales of heavy particles, where, $$M_R\approx M_{W_R}\approx M_{{\Delta_L}^{++}}\approx M_{{\Delta_R}^{++}}\approx TeV$$.
Under these assumptions, the amplitude for the light-heavy mixing contribution which is proportional to $\frac{{m_D}^2}{M_R}$ remains very small, since $m_\nu \approx \frac{{m_D}^2}{M_R} \approx (0.01-0.1)eV, m_D \approx (10^{5}-10^{6})eV$ which implies $\frac{m_D}{M_R}$ $\approx$ $(10^{-7}-10^{-6})eV$. Thus in our model, we ignore the contributions involving the light and heavy neutrino mixings.  

When NDBD is done in the framework of LRSM, the standard light neutrino contribution is given by,
\begin{equation}
	\label{e:45}
	{m_v^{eff}}= {U^2_{Li}}{m_i}
\end{equation}
where, $U_{Li}$ are the elements of the first row of the neutrino mixing matrix $U_{PMNS}$, in which the elements are dependent on known mixing angles $\theta_{13}$ , $\theta_{12}$ and the Majorana phases $\kappa$ and $\eta$. The $U_{PMNS}$ matrix is given by,
\begin{equation}
	\label{e:46}
	U_{PMNS} = \begin{pmatrix}
		c_{12}c_{13} & s_{12}c_{13} & s_{13}e^{-i\delta}\\
		-c_{23}s_{12} - s_{23}s_{13}c_{12}e^{i\delta} & -c_{23}c_{12} - s_{23}s_{12}s_{13}e^{i\delta} & s_{23}c_{13}\\
		s_{23}s_{12} - c_{23}s_{13}c_{12}e^{i\delta} & -s_{23}c_{12} - c_{23}s_{13}s_{12}e^{i\delta} & c_{23}c_{13}
	\end{pmatrix}P
\end{equation}
where, $P = diag(1,e^{i\kappa}, e^{i\eta})$. So the effective mass can be parametrized in terms of the elements of the diagonalizing matrix and the eigenvalues as,
\begin{equation}
	\label{e:47}
	{m_v^{eff}} = m_1c_{12}^{2}c_{13}^{2} + m_2s_{12}^{2}c_{13}^{2}{e^{2i\kappa}} + m_3s_{13}^{2}{e^{2i\eta}}.
\end{equation}

\begin{center}
	\section{Numerical Analysis and Results}
\end{center}

In our present work, we have modified left-right symmetric model by incorporating $A_4$ modular symmetry for both type-I and type-II dominances. As we are using modular symmetry, the Yukawa couplings are expressed as expansions of $q$ as shown in equations \eqref{e:22},\eqref{e:23} and \eqref{e:24}. In our model, the value of $q$ is found to be of the order of $10^{-1}$. The aboslute value of the modulus should however be greater than 1.\\ 
\begin{equation}
	\label{e:48}
	\tau = Re(\tau) + Im(\tau)
\end{equation}
\begin{table}[H]
	\begin{center}
		\begin{tabular}{|c|c|c|c|}
			\hline
			& $Re(\tau)$ & $Im(\tau)$ & $|\tau|$\\
			\hline
			Range & [0.715,0.789] & [0.8,0.9] & [1.073,1.197]\\
			\hline
		\end{tabular}
		\caption{\label{tab:Table 4} Range of values corresponding to real and imaginary parts of the modulus $\tau$.}
	\end{center}
\end{table}
\begin{table}[H]
	\begin{center}
		\begin{tabular}{|c|c|c|}
			\hline
			Yukawa couplings & Normal Hierarchy & Inverted hierarchy\\
			\hline
			$Y_1 (min)$ & $1.29155 \times 10^{-9}$ & $1.32276$ $\times$ $10^{-7}$\\
			$Y_1 (max)$ & $8.22986$ $\times$ $10^{-7}$ & $9.21382$ $\times$ $10^{-7}$\\
			\hline
			$Y_2 (min)$ & $3.02229$ $\times$ $10^{-9}$ & $2.42826$ $\times$ $10^{-9}$\\
			$Y_2 (max)$ & $1.32006$ $\times$ $10^{-6}$ & $1.45952$ $\times$ $10^{-6}$\\
			\hline
			$Y_3 (min)$ & $3.39287$ $\times$ $10^{-9}$ & $3.759$ $\times$ $10^{-9}$\\
			$Y_3 (max)$ & $1.3165$ $\times$ $10^{-6}$ & $1.50082$ $\times$ $10^{-6}$\\
			\hline
		\end{tabular}
		\caption{\label{tab:Table 4} Range of Yukawa couplings $(Y_1,Y_2,Y_3)$ for both normal and inverted hierarchy.}
	\end{center}
\end{table}
\vspace{-1cm}
From table (V), it is seen that the absolute value of the modulus is greater than unity, which is the expected result. The range of the Yukawa couplings for our model is shown in the table above. It is seen from the table that the minimum value of the Yukawa coupling $Y_1$ is in the scale of $10^{-10}$ for normal hierarchy while for inverted hierarchy it is in the scale of $10^{-7}$. However, for the maximum of $Y_1$ both the orderings are in the same scale. For $Y_2$ and $Y_3$, the minimum and maximum values for both normal and inverted hierarchies are within the same scale. We have plotted the effective masses against the Yukawa couplings $(Y_1,Y_2,Y_3)$ and it was found to be well within the bound set by experiments.\\
As shown both for type-I and type-II dominances, we have plotted the absolute values of the Yukawa couplings against the sum of the neutrino masses. The range of the values for sum of neutrino masses for both the cases are given as under,

 \begin{table}[H]
 	\begin{center}
 		\begin{tabular}{|c|c|c|}
 			\hline
 			$\sum m_\nu$ & Normal Hierarchy & Inverted hierarchy\\
 			\hline
 			$Type-I (min)$ & $0.000980556$ & $0.000437758$\\
 			$Type-I (max)$ & $0.177296$ & $0.186377$\\
 			\hline
 			$Type-II (min)$ & $0.000219304$ & $0.000035$\\
 			$Type-II (max)$ & $0.0200981$ & $0.0203081$\\
 			\hline
 		\end{tabular}
 		\caption{\label{tab:Table 4} Range of values for sum of neutrino masses for type-I and type-II dominances for both normal and inverted hierarchy.}
 	\end{center}
 \end{table}
\subsection{Standard Light Neutrino Contribution to $0\nu\beta\beta$}
As mentioned above, in the standard light neutrino contribution to $0\nu\beta\beta$, the intermediate particles are the $W_L$ bosons and light neutrino. The effective mass for the contribution is given by equation (5.2). Simplifying for the respective elements of $U_{Li}$ and $m_i$, the value of the effective mass is obtained in terms of the modular forms $(Y_1,Y_2,Y_3)$ as,
\begin{equation}
	\label{e:49}
	m_\nu^{eff} = m_1^{eff} + m_2^{eff} +m_3^{eff}
\end{equation}
where,
\begin{equation*}
	m_1^{eff} = \frac{\nu(Y_2-Y_3)^2 (Y_1+Y_2+Y_3)}{\nu_R(Y_1-Y_3)^2}
\end{equation*}
\begin{equation*}
	m_2^{eff} = \frac{\nu^2(Y_1-Y_2)^2(Y_1+Y_2+Y_3-\sqrt{3}\sqrt{3Y_1^2-2Y_1Y_2+3Y_2^2-2Y_1Y_3-2Y_2Y_3+3Y_3^2})}{2\nu_R(Y_1-Y_3)^2}
\end{equation*}
\begin{equation*}
	m_3^{eff} = \frac{\nu^2(Y_1+Y_2+Y_3-\sqrt{3}\sqrt{3Y_1^2-2Y_1Y_2+3Y_2^2-2Y_1Y_3-2Y_2Y_3+3Y_3^2})}{2\nu_R}
\end{equation*}
for type-I dominance, and the plots are shown as,
\begin{figure}[H]
	\centering
	\captionsetup{justification=centering}
	\includegraphics[scale=0.4]{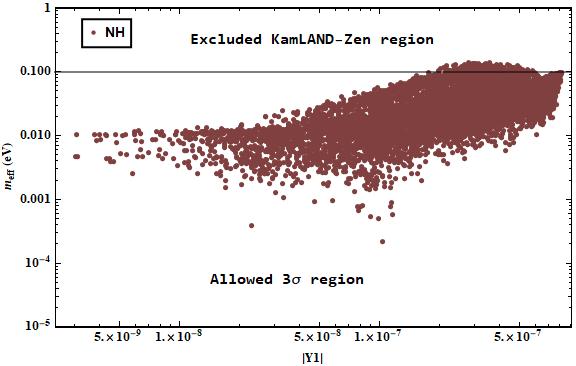}
	\includegraphics[scale=0.4]{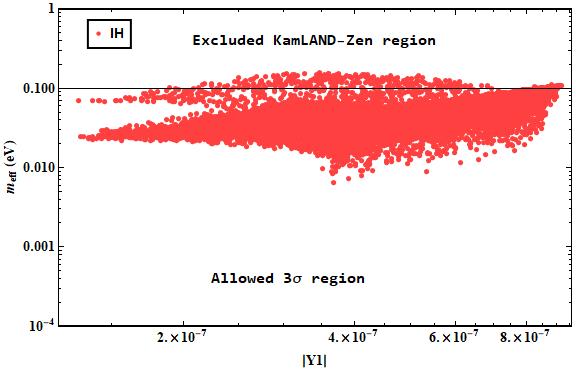}\\
	\caption{Variation of $|Y_1|$ with effective neutrino mass for standard light neutrino contribution.}
\end{figure}
\begin{figure}[H]
	\centering
	\captionsetup{justification=centering}
	\includegraphics[scale=0.4]{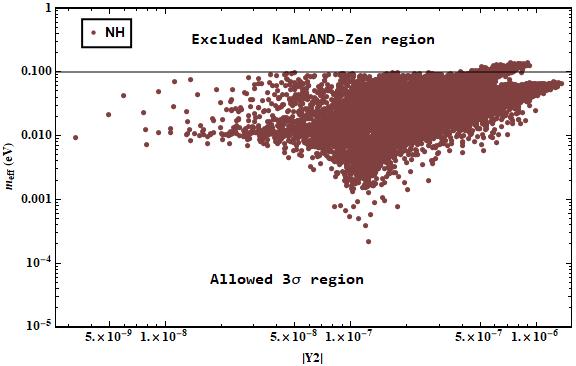}
	\includegraphics[scale=0.4]{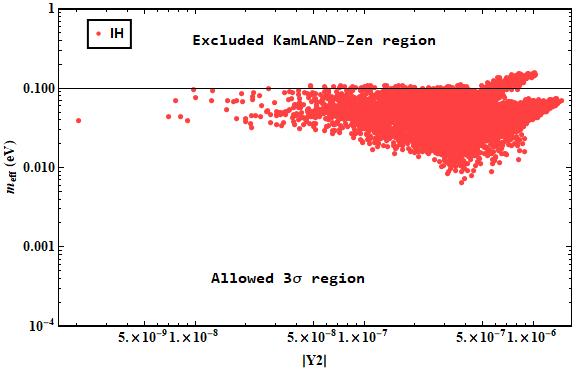}\\
	\caption{Variation of $|Y_2|$ with effective neutrino mass for standard light neutrino contribution.}
\end{figure}
\begin{figure}[H]
	\centering
	\captionsetup{justification=centering}
	\includegraphics[scale=0.4]{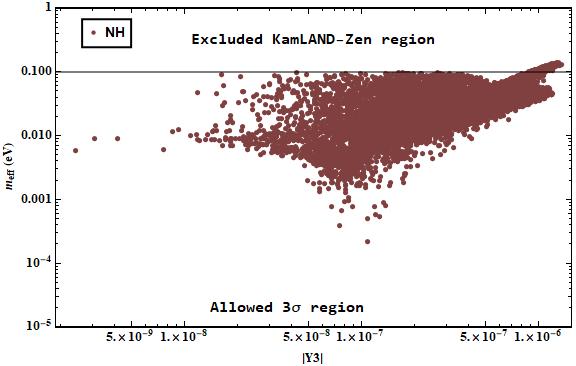}
	\includegraphics[scale=0.4]{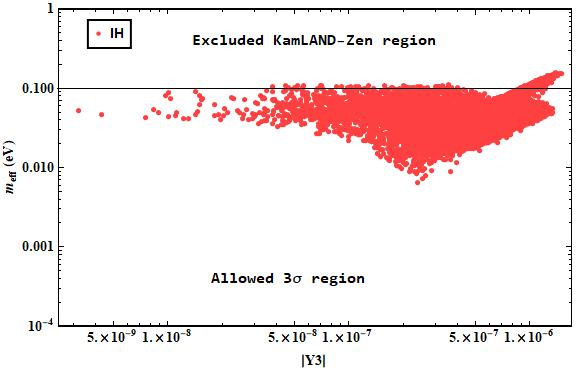}\\
	\caption{Variation of $|Y_3|$ with effective neutrino mass for standard light neutrino contribution.}
\end{figure}
For type-II dominance, we have
\begin{equation*}
	m_1^{eff} = -\frac{\nu_L(-Y_2+Y_3)(Y_1+Y_2+Y_3)}{Y_1-Y_2}
\end{equation*}
\begin{equation*}
	m_2^{eff} = -\frac{\nu_L (Y_1 - Y_3)(Y_1 + Y_2 + Y_3 -\sqrt{3}\sqrt{3 Y_1^2 - 2 Y_1 Y_2 + 3 Y_2^2 - 2 Y_1 Y_3 - 2 Y_2 Y_3 + 3 Y_3^2})}{2(Y_1-Y_2)}
\end{equation*}
\begin{equation*}
	m_3^{eff} =\frac{\nu_L(Y_1 - Y_3)(Y_1 + Y_2 + Y_3 +\sqrt{3}\sqrt{3 Y_1^2 - 2 Y_1 Y_2 + 3 Y_2^2 - 2 Y_1 Y_3 - 2 Y_2 Y_3 + 3 Y_3^2})}{2}
\end{equation*}
\begin{figure}[H]
	\centering
	\captionsetup{justification=centering}
	\includegraphics[scale=0.4]{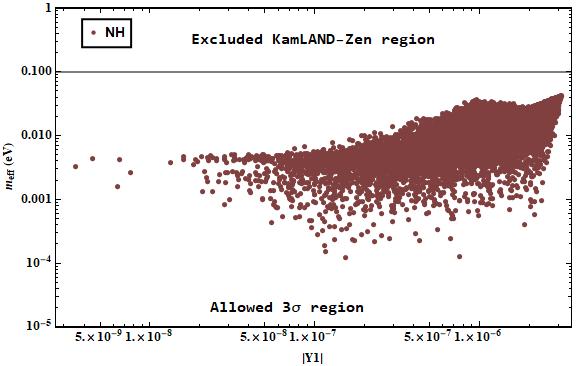}
	\includegraphics[scale=0.4]{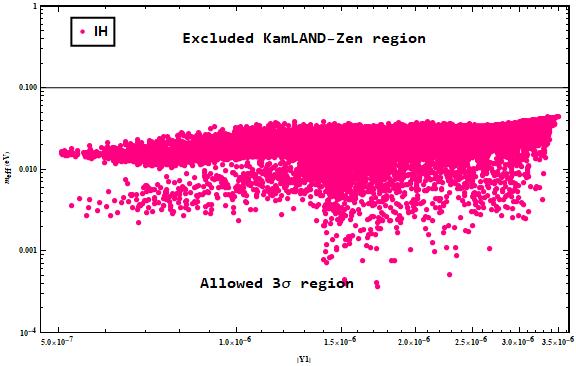}\\
	\caption{Variation of $|Y_1|$ with effective neutrino mass for standard light neutrino contribution.}
\end{figure}
\begin{figure}[H]
	\centering
	\captionsetup{justification=centering}
	\includegraphics[scale=0.4]{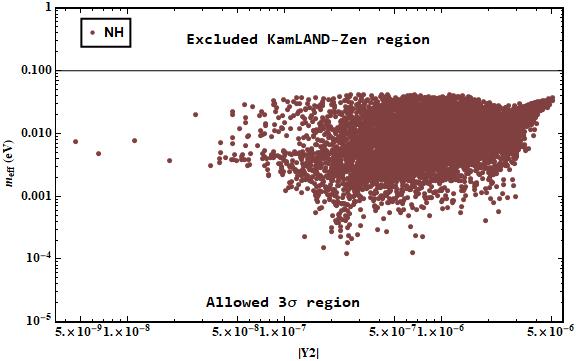}
	\includegraphics[scale=0.4]{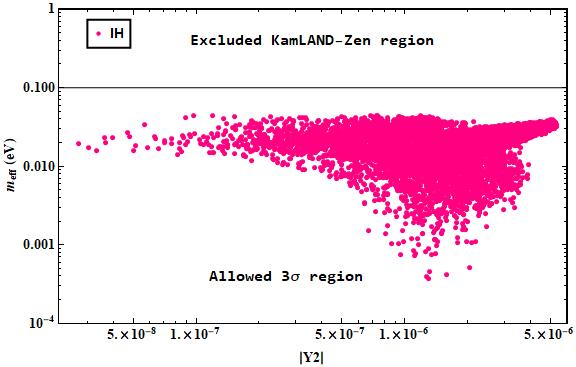}\\
	\caption{Variation of $|Y_2|$ with effective neutrino mass for standard light neutrino contribution.}
\end{figure}
\begin{figure}[H]
	\centering
	\captionsetup{justification=centering}
	\includegraphics[scale=0.4]{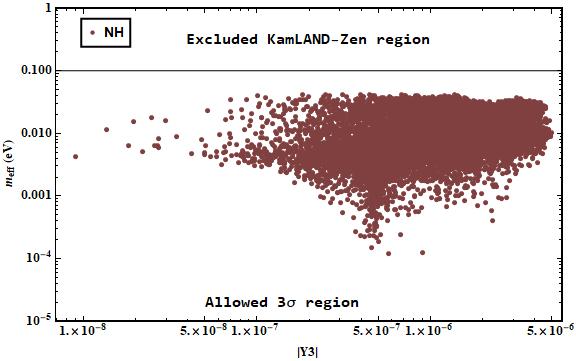}
	\includegraphics[scale=0.4]{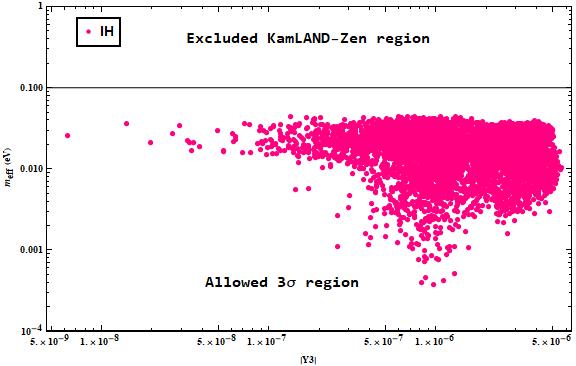}\\
	\caption{Variation of $|Y_3|$ with effective neutrino mass for standard light neutrino contribution.}
\end{figure}
\subsection{Heavy Right-Handed Neutrino contribution to $0\nu\beta\beta$}
In our work, we have considered contributions of heavy right-handed neutrino and scalar Higgs triplet to NDBD. The effective mass for heavy right-handed neutrino is given by,
\begin{equation}
	\label{e:50}
	m_R^{eff} = p^2\Biggl(\frac{M_{W_L}^4}{M_{W_R}^4}\Biggl)\Biggl(\frac{U_{Rei}^{{*}^2}}{M_i}\Biggl)
\end{equation}
where, $p^2$ is the typical momentum exchange of the process. As it is known that TeV scale LRSM plays a very important role in the process of neutrinoless double beta decay (0$\nu\beta\beta$), we have considered the values as $M_{W_R} = 10TeV$ , $M_{W_L} = 80GeV$ , $M_{\Delta_R} \approx 3TeV$  and after calculation, the value for heavy right-handed neutrino is found to be in the scale of $TeV$. The allowed value of p is in the range $(100-200)MeV$ and so we consider, $p \approx 180 MeV$.
Thus, we get,
\begin{equation}
	\label{e:51}
	p^2\Biggl(\frac{M_{W_L}^4}{M_{W_R}^4}\Biggl) = 10^{10} eV^2
\end{equation}
where, $U_{Rei}$ refers to the first row elements of the diagonalizing matrix of the heavy Majorana mass matrix and $M_i$ are its eigenvalues.
The effective mass corresponding to the heavy right-handed neutrino can be expressed in terms of the modular forms as,
\begin{equation}
	\label{e:52}
	m_{eff}^R = 10^{10}(m_{eff}^{R_{1}} + m_{eff}^{R_{2}} + m_{eff}^{R_{3}})
\end{equation} 
where,
\begin{equation*}
	m_{eff}^{R_{1}} = \frac{2}{\nu_R (Y_1 + Y_2 + Y_3 + \sqrt{3}\sqrt{3 Y_1^2 - 2 Y_1 Y_2 + 3 Y_2^2 - 2 Y_1 Y_3 - 2 Y_2 Y_3 + 3 Y_3^2})}
\end{equation*}
\begin{equation*}
	m_{eff}^{R_{2}} = \frac{2(Y_1^{*} - Y_3^{*})^{2}}{\nu_R(Y_1 + Y_2 + Y_3 -\sqrt{3}\sqrt{3 Y_1^2 - 2 Y_1 Y_2 + 3 Y_2^2 - 2 Y_1 Y_3 - 2 Y_2 Y_3 + 3 Y_3^2})(Y_1^{*} - Y_2^{*})^{2}}
\end{equation*}
\begin{equation*}
	m_{eff}^{R_{3}} = \frac{(-Y_2^{*} + Y_3^{*})^{2}}{\nu_R(Y_1 + Y_2 + Y_3)(Y_1^{*} - Y_2^{*})^{2}}
\end{equation*}
The total effective mass is also calculated for the standard light and right-handed heavy neutrino contribution, given by,
\begin{equation}
	\label{e:53}
	|m_\nu^{{eff}^{total}}| = |m_\nu^{eff} + m_{eff}^R|
\end{equation}
which can be obtained in terms of the modular forms as a summation of the above mentioned terms.
\begin{figure}[H]
	\centering
	\captionsetup{justification=centering}
	\includegraphics[scale=0.4]{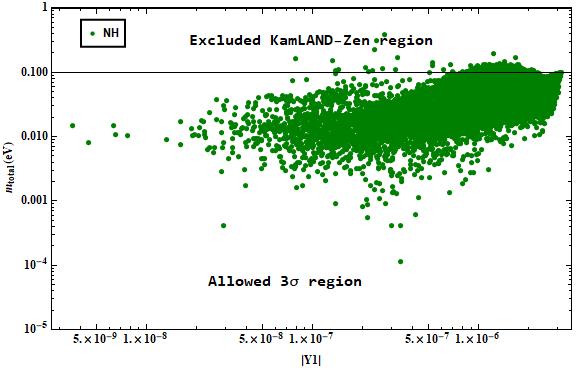}
	\includegraphics[scale=0.4]{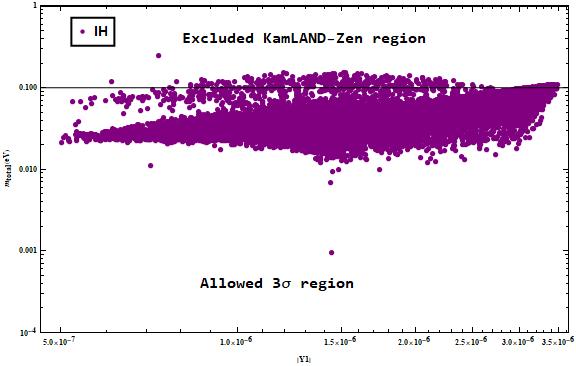}\\
	\caption{Variation of $|Y_1|$ with total effective neutrino mass.}
\end{figure}
\begin{figure}[H]
	\centering
	\captionsetup{justification=centering}
	\includegraphics[scale=0.4]{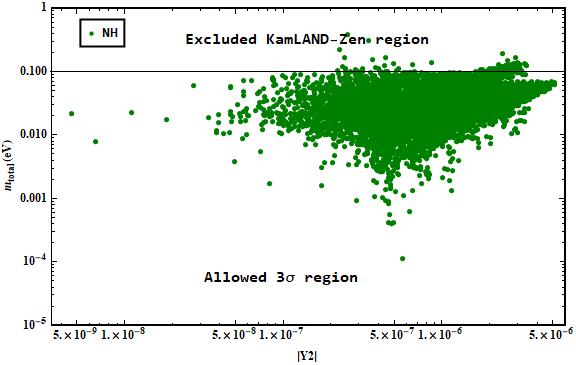}
	\includegraphics[scale=0.4]{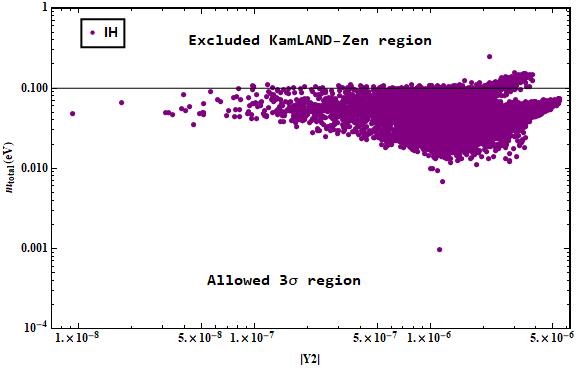}\\
	\caption{Variation of $|Y_2|$ with total effective neutrino mass.}
\end{figure}
\begin{figure}[H]
	\centering
	\captionsetup{justification=centering}
	\includegraphics[scale=0.4]{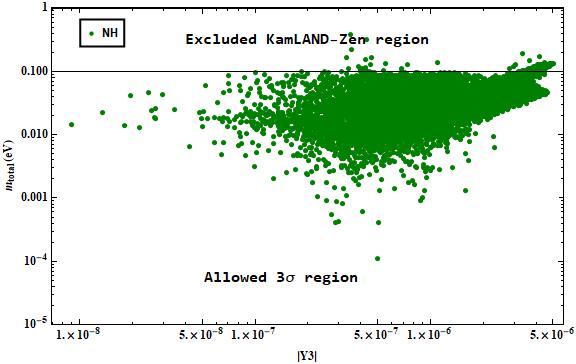}
	\includegraphics[scale=0.4]{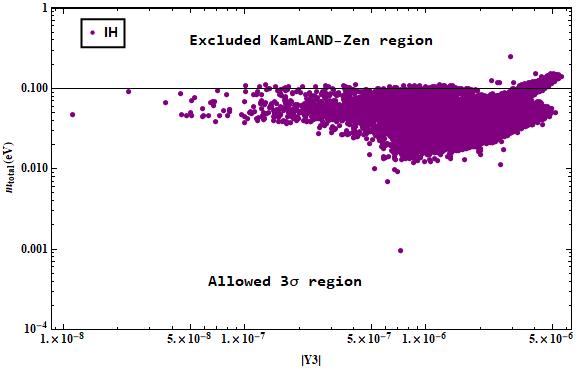}\\
	\caption{Variation of $|Y_3|$ with total effective neutrino mass.}
\end{figure}
The plots above are for type-I dominance.
\begin{figure}[H]
	\centering
	\captionsetup{justification=centering}
	\includegraphics[scale=0.4]{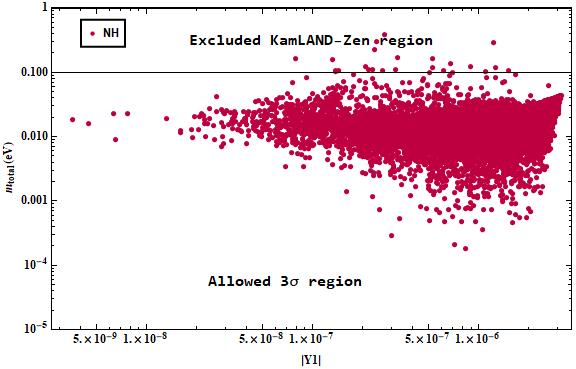}
	\includegraphics[scale=0.4]{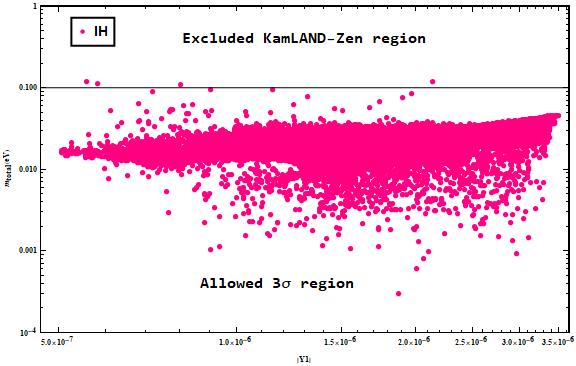}\\
	\caption{Variation of $|Y_1|$ with total effective neutrino mass.}
\end{figure}
\begin{figure}[H]
	\centering
	\captionsetup{justification=centering}
	\includegraphics[scale=0.4]{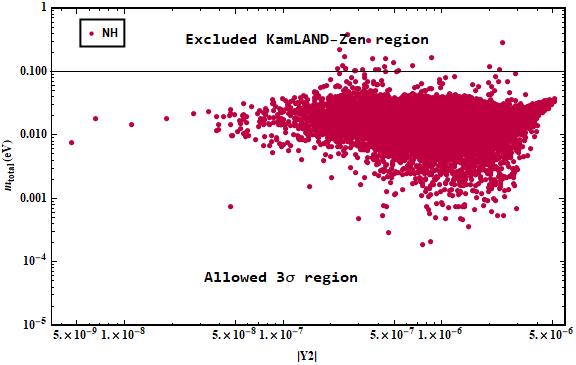}
	\includegraphics[scale=0.4]{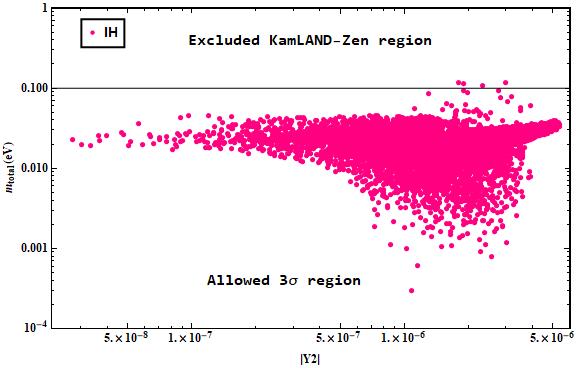}\\
	\caption{Variation of $|Y_2|$ with total effective neutrino mass.}
\end{figure}
\begin{figure}[H]
	\centering
	\captionsetup{justification=centering}
	\includegraphics[scale=0.4]{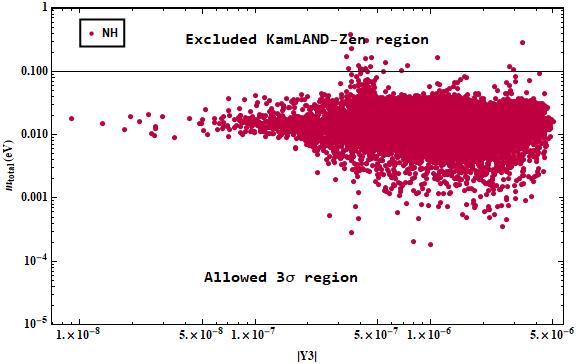}
	\includegraphics[scale=0.4]{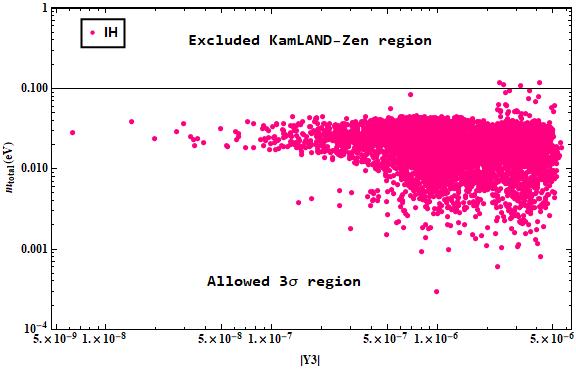}\\
	\caption{Variation of $|Y_3|$ with total effective neutrino mass.}
\end{figure}
The figures above are the plots for type-II dominance.
\subsection{Scalar Triplet contribution to $0\nu\beta\beta$}
The magnitude of $\Delta_R$ contribution is controlled by the factor $\frac{M_i}{M_{\Delta_R}}$ \cite{Chakrabortty:2012mh}. However, scalar triplet contribution is not included in the total contribution under the assumption $\frac{M_i}{M_{\Delta_R}}<0.1$. But, some the mixing parameters in the large part of the parameter space may result in a higher $\frac{M_i}{M_{\Delta_R}}$ ratio and in such cases we will have to include it in the total contribution. The impact of this contribution here is studied in the limit, $M_{\Delta_R} \approxeq M_{heaviest}$.\\
The effective mass for scalar triplet contribution is given as,
\begin{equation}
	\label{e:54}
	|m_\Delta^{eff}| = |p^2 \frac{M_{W_L}^4}{M_{W_R}^4} \frac{U_{Rei}^2M_i}{M_{\Delta_R}^2}|
\end{equation}
The value of the mass for the right-handed scalar triplet is taken as, $M_{\Delta_R} = 3TeV$. So, the value of the coefficient results as,
\begin{equation}
	\label{e:55}
	p^2 \frac{M_{W_L}^4}{M_{W_R}^4} \frac{1}{M_{\Delta_R}^2} = \frac{10^{10}}{9 \times 10^{24}}
\end{equation} 
In terms of modular forms, the effective scalar mass can be expressed as,
\begin{equation}
	\label{e:56}
	m_{eff}^{\Delta_R} = m_{eff_1}^{\Delta_R}+m_{eff_2}^{\Delta_R}+m_{eff_3}^{\Delta_R}
\end{equation}
where,
\begin{equation*}
	m_{eff_1}^{\Delta_R} = \frac{\nu_R(Y_2+Y_3)^2(Y_1+Y_2Y_3)}{(Y_1-Y_2)^2}
\end{equation*}
\begin{equation*}
	m_{eff_2}^{\Delta_R} = \frac{\nu_R(Y_1-Y_3)^2(Y_1 + Y_2 + Y_3 -\sqrt{3}\sqrt{3 Y_1^2 - 2 Y_1 Y_2 + 3 Y_2^2 - 2 Y_1 Y_3 - 2 Y_2 Y_3 + 3 Y_3^2})}{2(Y_1-Y_2)^2}
\end{equation*}
\begin{equation*}
	m_{eff_3}^{\Delta_R} = \frac{\nu_R(Y_1 + Y_2 + Y_3 +\sqrt{3}\sqrt{3 Y_1^2 - 2 Y_1 Y_2 + 3 Y_2^2 - 2 Y_1 Y_3 - 2 Y_2 Y_3 + 3 Y_3^2})}{2}
\end{equation*}
The plots are shown as under.
\begin{figure}[H]
	\centering
	\captionsetup{justification=centering}
	\includegraphics[scale=0.4]{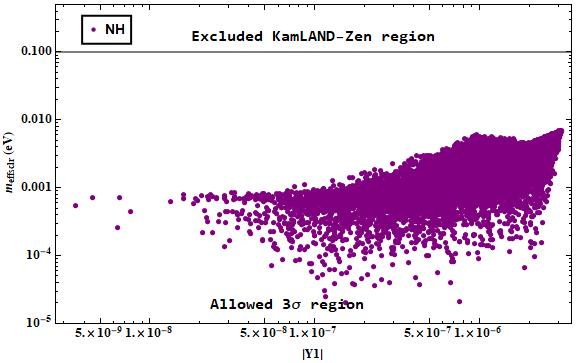}
	\includegraphics[scale=0.4]{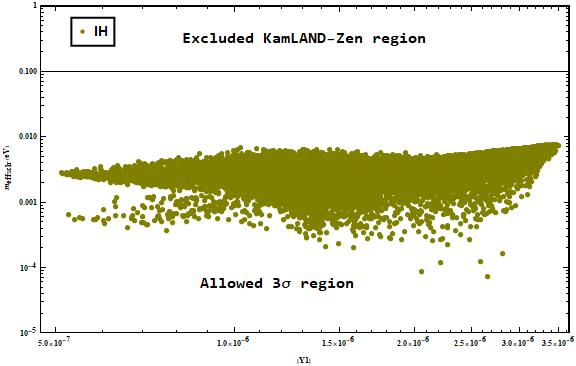}\\
	\caption{Variation of $|Y_1|$ with effective neutrino mass for scalar triplet contribution.}
\end{figure}
\begin{figure}[H]
	\centering
	\captionsetup{justification=centering}
	\includegraphics[scale=0.4]{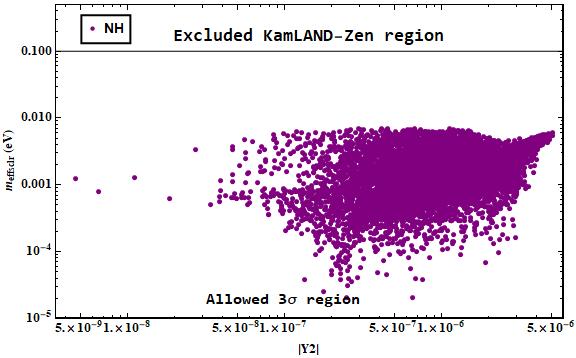}
	\includegraphics[scale=0.4]{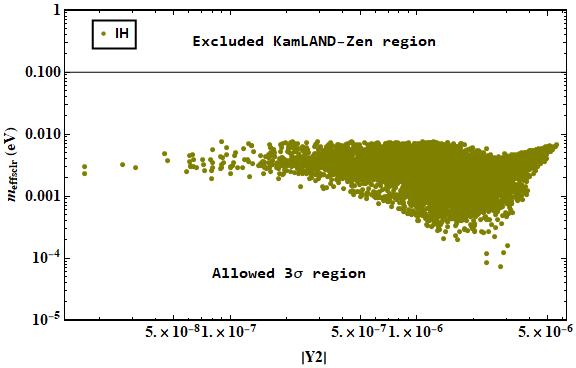}\\
	\caption{Variation of $|Y_2|$ with effective neutrino mass for scalar triplet contribution.}
\end{figure}
\newpage
\begin{figure}[H]
	\centering
	\captionsetup{justification=centering}
	\includegraphics[scale=0.4]{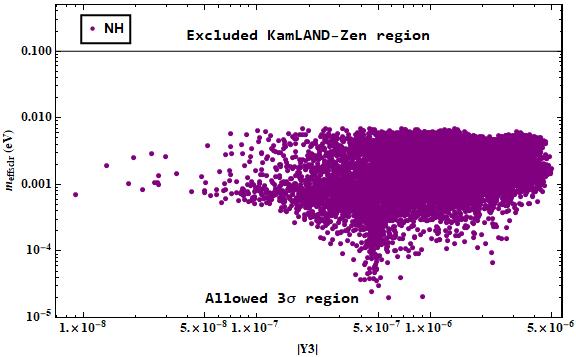}
	\includegraphics[scale=0.4]{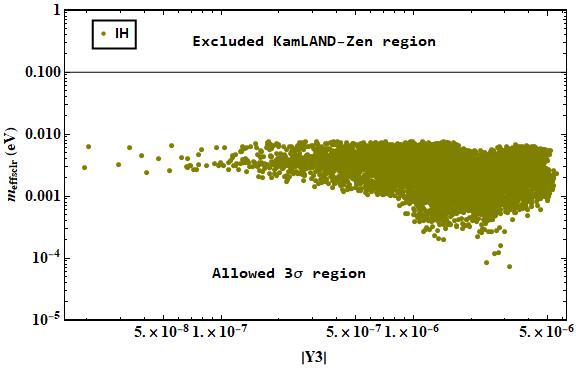}\\
	\caption{Variation of $|Y_3|$ with effective neutrino mass for scalar triplet contribution.}
\end{figure}
\begin{center}
	\section{Conclusion}
\end{center}

The discovery of neutrino oscillations paved the gateway for physics beyond the Standard Model. In our paper, we have realized LRSM with the help of modular $A_4$ symmetry for both type-I and type-II dominance. Using modular symmetry provides the advantage of using no extra particles called 'flavons'. The Yukawa couplings are represented as modular forms expressed as expansions of $q$. The values of the Yukawa couplings $(Y_1,Y_2,Y_3)$ are calculated using 'Mathematica'. The mass matrices are then determined using the multiplication rules for $A_4$ group stated in the Appendix. The Majorana mass matrix is found to be symmetric and under the considered basis, the charged lepton mass matrix is also diagonal. We have expressed the light neutrino and heavy right-handed neutrino mass matrix in terms of the modular forms. We have also studied briefly the contributions of $0\nu\beta\beta$ in LRSM. The effective masses corresponding to standard light neutrino contribution, right-handed contribution and scalar triplet contributions are determined in terms of $(Y_1,Y_2,Y_3)$ and we have plotted the effective mass corresponding to the considered contributions against the Yukawa couplings. To summarize our work, some results are stated as under,
\begin{itemize}
	\item The absolute value of the modulus was found to be within the range $1.073$ to $1.197$, which is greater than unity, that is the desired result.
	\item The Yukawa couplings, expressed in terms of modular forms ranges from $10^{-9}$ to $10^{-6}$.
	\item The sum of the neutrino masses for type-I dominance ranges from
	the order of $10^{-4}$ to $10^{-1}$ for both normal and inverted hierarchy.
	\item The sum of the neutrino masses for type-II dominance ranges from
	the order of $10^{-4}$ to $10^{-2}$ for both normal and inverted hierarchy.
\end{itemize}
The effective masses for the $0\nu\beta\beta$ contributions are calculated and by determining their relations with the modular forms, we have plotted the effective masses with the three Yukawa couplings and it has been found that the values for the effective mass corresponding to each contribution is well within the experimental bounds, which infact makes us clearly state that the building of the model with modular symmetry is advantageous to that of flavor symmetries. In this model, we have not used any extra particles and the analysis has been done taking into consideration the calculated and computed values for the model parameters and the results are found to be satisfactory, so it can be stated that the Left-Right Symmetric Model can be constructed with modular symmetry while satisfying the experimental bounds on the desired parameters.

\begin{center}
	\section{Appendix A}
\end{center}
Let us consider the Higgs potential of our model that has quadratic and quartic coupling terms given by \cite{Luo:2008rs},
\begin{equation*}
	V_{\phi,\Delta_L,\Delta_R} = -\mu_{ij}^2 Tr[\phi_i^\dagger \phi_j] +\lambda_{ijkl} Tr[\phi_i^\dagger\phi_j]Tr[\phi_k^\dagger \phi_l]+ \lambda_{ijkl}^{'} Tr[\phi_i^\dagger\phi_j\phi_k^\dagger \phi_l] -\mu_{ij}^2 Tr[\Delta_L^\dagger \Delta_L + \Delta_R^\dagger \Delta_R] 
\end{equation*}
\begin{equation*}
	\rho_1 [(Tr[\Delta_L^\dagger \Delta_L])^2 + (Tr[\Delta_L^\dagger \Delta_L])^2] + \rho_2 (Tr[\Delta_L^\dagger \Delta_L \Delta_L^\dagger \Delta_L] + Tr[\Delta_R^\dagger \Delta_R \Delta_R^\dagger \Delta_R]) +\rho_3 Tr[\Delta_L^\dagger \Delta_L \Delta_R^\dagger \Delta_R] + 
\end{equation*}
\begin{equation}
	\label{e:57}
	\alpha_{ij} Tr[\phi_i^\dagger \phi_j](Tr[\Delta_L^\dagger \Delta_L] + Tr[\Delta_R^\dagger \Delta_R]) + \beta_{ij}(Tr[\Delta_L^\dagger \Delta_L \phi_i \phi_j^\dagger] + Tr[\Delta_R^\dagger \Delta_R \phi_i \phi_j^\dagger]) + \gamma_{ij}(Tr[\Delta_L^\dagger \phi_i \Delta_R \phi_j^\dagger] +h.c)
\end{equation}
where, i,j,k,l runs from 1 to 2 with $\phi_1 = \phi$ and $\phi_2 = \tilde{\phi}$.
As mentioned above after SSB, the scalar sector obtains VEV. So after the substitution of the respective VEVs and determining the traces, so after simplification the potential can be written as,
\begin{equation}
	\label{e:58}
	V = -\mu^2 (v_L^2 + v_R^2 ) + \frac{\rho}{4} (v_L^4 + v_R^4) + \frac{\rho'}{2} + \frac{\alpha}{2}(v_L^2 + v_R^2) k_1^2 + \gamma v_L v_R k^2
\end{equation}
where, we have used the approximation $k' << k$, and $\rho' = 2\rho_3$.
Our minimization conditions are, $\frac{\delta V}{\delta v_L} = \frac{\delta V}{\delta v_R} = \frac{\delta V}{\delta k} = \frac{\delta V}{\delta k'} = 0$

Therefore, we get,
\begin{equation}
	\frac{\delta V}{\delta v_L} = -2\mu^2 v_L + \rho v_L^3 + \rho' v_L k^2 + \gamma v_R k^2
\end{equation}
Here, it is evident that the Majorana mass of the left-handed neutrino $M_{LL}$ is dependent on the vev $v_L$ as already defined above.
Again, we have
\begin{equation}
	\frac{\delta V}{\delta v_R} = -2\mu^2 v_R + \rho v_R^3 + \rho' v_R k^2 + \gamma v_L k^2
\end{equation}
So, the right handed Majorana mass $M_{RR}$ is dependent on the vev $v_R$. Similarly, the calculations for the same can be carried out and it can be found out the Dirac mass term $M_D$ can be expressed in terms of the vev for the Higgs bidoublet as also defined previously.\\
Now, we are to determine a relation between the VEVs for the scalars and so after using the minimization conditions and simplifying the equations, we come to a relation given by,
\begin{equation}
	v_L v_R = \frac{\gamma}{\xi} k
\end{equation}
where, $\xi = \rho - \rho'$.\\ 
The neutrino mass for LRSM is given as a summation of the type-I and type-II term as already mentioned above. So, in the approximation that $k'<< k$, and if we consider that our Yukawa coupling $Y^l$ corresponding to the neutrino masses is $y_D$ and the coupling $\widetilde{Y^l}$ for the charged fermion masses is denoted by $y_L$, so considering $y_D k >> y_l k'$  we can write,
\begin{equation}
	M_\nu = \frac{k^2}{v_R} y_D f_R^{-1} y_D^T + f_L v_L
\end{equation}
Since, for due to left-right symmetry, we can consider $f_L = f_R =f$, so the above equation can be written as,
\begin{equation}
	M_\nu = \frac{k^2}{v_R} y_D f^{-1} y_D^T + f v_L
\end{equation}
So, from this equation we can come to a relation given by,
\begin{equation}
	M_\nu = (f\frac{\gamma}{\xi} + y_D f^{-1} y_D^T)\frac{k^2}{v_R}
\end{equation}
Here, we can consider two situations, namely
\begin{itemize}
	\item If $f(\frac{\gamma}{\xi}) << y_D f^{-1}y_D^T$, the light neutrino mass is given by the type-I term $M_D M_{RR}^{-1}M_D^T$. That is, here type-I is dominant and the light neutrino mass is from the suppression of heavy $\nu_R$.
	\item If $f(\frac{\gamma}{\xi}) >> y_D f^{-1}y_D^T$, the light neutrino mass is given by the type-II term $f v_L$. That is, in this case type-II mass term is dominant and the light neutrino mass is because of the tiny value of $\nu_L$.
\end{itemize}
\begin{center}
	\section{Appendix B}
\end{center}
\subsection*{Properties of $A_4$ group}
$A_4$ is a non-abelian discrete symmetry group which represents even permuatations of four objects. It has four irreducible representations, three out of which are singlets $(1,1',1'')$ and one triplet $3$ ($3_A$ represents the anti-symmetric part and $3_S$ the symmetric part). Products of the singlets and triplets are given by,
\begin{center}
	\begin{equation*}
		1 \otimes 1 = 1
	\end{equation*}
\end{center}
\begin{center}
	\begin{equation*}
		1' \otimes 1' = 1''
	\end{equation*}
\end{center}
\begin{center}
	\begin{equation*}
		1' \otimes 1'' = 1
	\end{equation*}
\end{center}
\begin{center}
	\begin{equation*}
		1'' \otimes 1'' = 1'
	\end{equation*}
\end{center}
\begin{center}
	\begin{equation*}
		3 \otimes 3 = 1 \oplus 1' \oplus 1'' \oplus 3_A \oplus 3_S
	\end{equation*}
\end{center}
If we have two triplets under $A_4$ say, $(a_1,a_2,a_3)$ and $(b_1,b_2,b_3)$ , then their multiplication rules are given by,
\begin{center}
	\begin{equation*}
		1 \approx a_1b_1 + a_2b_3 + a_3b_2
	\end{equation*}
	\begin{equation*}
		1' \approx a_3b_3 + a_1b_2 + a_2b_1
	\end{equation*}
	\begin{equation*}
		1'' \approx a_2b_2 + a_3b_1 + a_1b_2
	\end{equation*}
	\begin{equation*}
		3_S \approx \begin{pmatrix}
			2a_1b_1-a_2b_3-a_3b_2 \\
			2a_3b_3-a_1b_2-a_2b_1 \\
			2a_2b_2-a_1b_3-a_3b_1
		\end{pmatrix}
	\end{equation*}
	\begin{equation*}
		3_A \approx \begin{pmatrix}
			a_2b_3-a_3b_2 \\
			a_1b_2-a_2b_1 \\
			a_3b_1-a_1b_3
		\end{pmatrix}
	\end{equation*}
\end{center}
\vspace{1cm}
\begin{center}
	\section{References}
\end{center}
\bibliographystyle{unsrt}
\bibliography{cite}

\end{document}